# Superbunched random fiber laser

*Qiang Ji*[1,2,†], *Yuxi Pang*[1,2,†], *Weiming Zhang*[1,2], *Xian Zhao*[1,2], *Zengguang Qin*[2,3], *Zhaojun Liu*[2,3,**], *and Yanping Xu*[1,2,*]

[1] *Center for Optics Research and Engineering, Shandong University, Qingdao 266237, China*

[2] *Key Laboratory of Laser and Infrared System of Ministry of Education, Shandong University, Qingdao 266237, China*

[3] *School of Information Science and Engineering, Shandong University, Qingdao 266237, China*

[†] *These authors contributed equally to this work.*

[**] *zhaojunliu@sdu.edu.cn*.

[*] *Correspondence and requests for materials should be addressed to Y. X. (yanpingxu@sdu.edu.cn).*

## Abstract

Photon superbunching, distinguished by second-order coherence values far exceeding the Gaussian thermal limit, represents a highly desirable resource for quantum optics and correlation-based imaging technologies. However, existing approaches typically rely on fragile experimental platforms, inefficient nonlinear conversion processes, or mechanically complex optical architectures. Here, we demonstrate a fully fiber-integrated superbunched random fiber laser (SRFL) in which intrinsic Rayleigh scattering cooperatively interacts with cascaded stimulated Brillouin scattering and quasi-phase-matched four-wave mixing to tailor extreme photon statistics. The SRFL generates a multi-wavelength comb, in which individual spectral components exhibit widely tunable photon bunching, with the second-order coherence $g^{(2)}(0)$ continuously controlled from ~1 to ~26 by tuning the pump power, spectral order and diffusion length. Moreover, we establish a direct correlation between photonic phase transitions (quantified by the Parisi overlap order parameter) and the emergence of superbunching, thereby bridging macroscopic disorder physics and microscopic photon statistics. Finally, we employ the superbunched emission for temporal ghost imaging, realizing high-fidelity temporal object reconstruction with a substantial reduction in required ensemble averaging. These findings validate random fiber lasers as a robust, scalable, and integrated platform for generating extreme photon statistics and unlock new avenues for correlation-enhanced photonic sensing and quantum optics investigations in complex photonic systems.

## Introduction

Random fiber lasers[1,2,3] (RFLs) provide a versatile platform for investigating light-matter interactions in disordered photonic media. The random distributed feedback resulting from Rayleigh scattering, combined with gain and nonlinear interactions, has enabled the observation of Anderson localization[4], optical rogue waves[5,6], replica symmetry breaking[7,8] and non-equilibrium photonic phase transitions[9] in one-dimensional fiber systems. These pioneering studies have elucidated the fundamental mechanisms by which disorder, distributed feedback and nonlinearity collectively govern the macroscopic optical behavior of RFLs[10,11]. However, the vast majority of reported work has focused on spectral, temporal or phase space features averaged over many experimental realizations[5-9,12-16]. The photon statistics and higher-order coherence properties of RFLs, particularly in operating regimes with strong disorder and cascaded nonlinear processes, remain largely unexplored[17,18]. This critical knowledge gap restricts the utility of RFLs as model systems for quantum optics studies in complex photonic media and constrains their practical impact on correlation-based optical sensing and imaging technologies[18-20].

The second-order coherence function $g^{(2)}(\tau)$ is a foundational quantity for characterizing photon statistics and quantum optical behavior[21-24]. Its zero-delay value $g^{(2)}(0)$ serves as a definitive metric to distinguish distinct light emission regimes: coherent light with $g^{(2)}(0)=1$, thermal or chaotic light with Gaussian field statistics for which $1< g^{(2)}(0)\leq 2$, and non-classical light fields such as anti-bunched single-photon streams with $g^{(2)}(0)<1$. Superbunching, defined as $g^{(2)}(0)$ value far exceeding the Gaussian-statistics thermal (chaotic) limit of 2, corresponds to extreme photon clustering and yields a substantial enhancement in correlation contrast. Such non-classical photon states are highly desirable for correlation-based sensing and ghost imaging applications, where the reconstruction efficiency and image visibility scale directly with the magnitude of $g^{(2)}(0)$[25-26]. Existing approaches to generating superbunched light have primarily relied on nonlinear optical interactions in atomic and semiconductor systems[27-33], or linear optical schemes based on multi-photon interference and mechanically modulated diffusers[34-39]. Yet these platforms suffer from inherent limitations including low conversion efficiency, narrow operating bandwidth, stringent environmental stability requirements, and limited scalability for large-scale photonic integration. RFLs are intrinsically fluctuation-rich optical systems: distributed Rayleigh scattering provides a dense ensemble of random feedback pathways, while stimulated Brillouin scattering (SBS) and quasi-phase-matched four-wave mixing (FWM) introduce strong nonlinear gain and efficient intermodal coupling. In such systems, disorder-induced linear diffusion and cascaded nonlinear dynamics act synergistically on the optical field, in principle enabling the generation of super-thermal intensity statistics that transcend the standard Gaussian thermal light limit. Despite this potential, the prevailing understanding of RFLs has emphasized their optical behavior as conventional or quasi-thermal light sources, and a clear physical connection between macroscopic phase transitions in disordered photonic media and the microscopic photon statistics of the emitted light has not yet been established. To date, a controlled realization of optical superbunching in a fully fiber-integrated random laser, alongside a quantitative mapping between disorder-driven photonic phases and photon correlation properties, has remained an unmet challenge.

In this work, we demonstrate a fully fiber-integrated superbunched random fiber laser (SRFL) in which intrinsic Rayleigh scattering synergizes with cascaded nonlinear interactions, specifically cascaded SBS (CSBS) and quasi-phase-matched FWM, to engineer extreme photon statistics in a deterministic manner. The SRFL emits a multi-wavelength comb, where individual spectral components exhibit widely and continuously tunable photon bunching with $g^{(2)}(0)$ values adjustable from ~1 up to ~26 by tailoring the pump power, spectral order and Rayleigh diffusion length. By analyzing the laser emission within a spin glass theoretical framework, we characterize photonic phase transitions using the Parisi overlap order parameter and identify a direct correspondence: the enhancement of superbunching coincides with the dissolution of the glassy photonic phase and the system's evolution toward a non-equilibrium paramagnetic regime. This finding establishes a rigorous quantitative link between macroscopic disorder physics and the microscopic photon statistics of the emitted light. Finally, we leverage the SRFL output to achieve high-fidelity temporal ghost imaging (TGI)[40,41,42] with a dramatic reduction in the required ensemble averaging. These results demonstrate that coherence engineering via controlled disorder and nonlinear coupling transforms the RFL into a highly efficient light source for correlation-enhanced optical sensing and imaging applications.

## Operating principle

The fundamental mechanism governing superbunching in the SRFL is illustrated in **Figs. 1a- 1c**. This extreme photon clustering phenomenon arises from the synergistic interplay between disorder-induced linear diffusion in the passive Rayleigh scattering fiber and cascaded nonlinear optical interactions in the active Brillouin gain fiber. As depicted in **Fig. 1a**, a continuous-wave pump beam is launched into the Brillouin gain fiber (BGF), where electrostriction-induced acoustic phonon excitation triggers stimulated Brillouin scattering (SBS) and generates the fundamental Stokes emission. This Stokes light subsequently propagates into the Rayleigh scattering fiber (RSF), undergoing multiple stochastic Rayleigh scattering events along its propagation path. A fraction of the backscattered Stokes light is retroreflected back into the BGF, counter-propagating with the incident pump beam to form a Rayleigh-scattering-mediated one-dimensional diffusion process (**Fig. 1c**), which provides robust distributed feedback and seeds subsequent Brillouin amplification in the gain fiber.

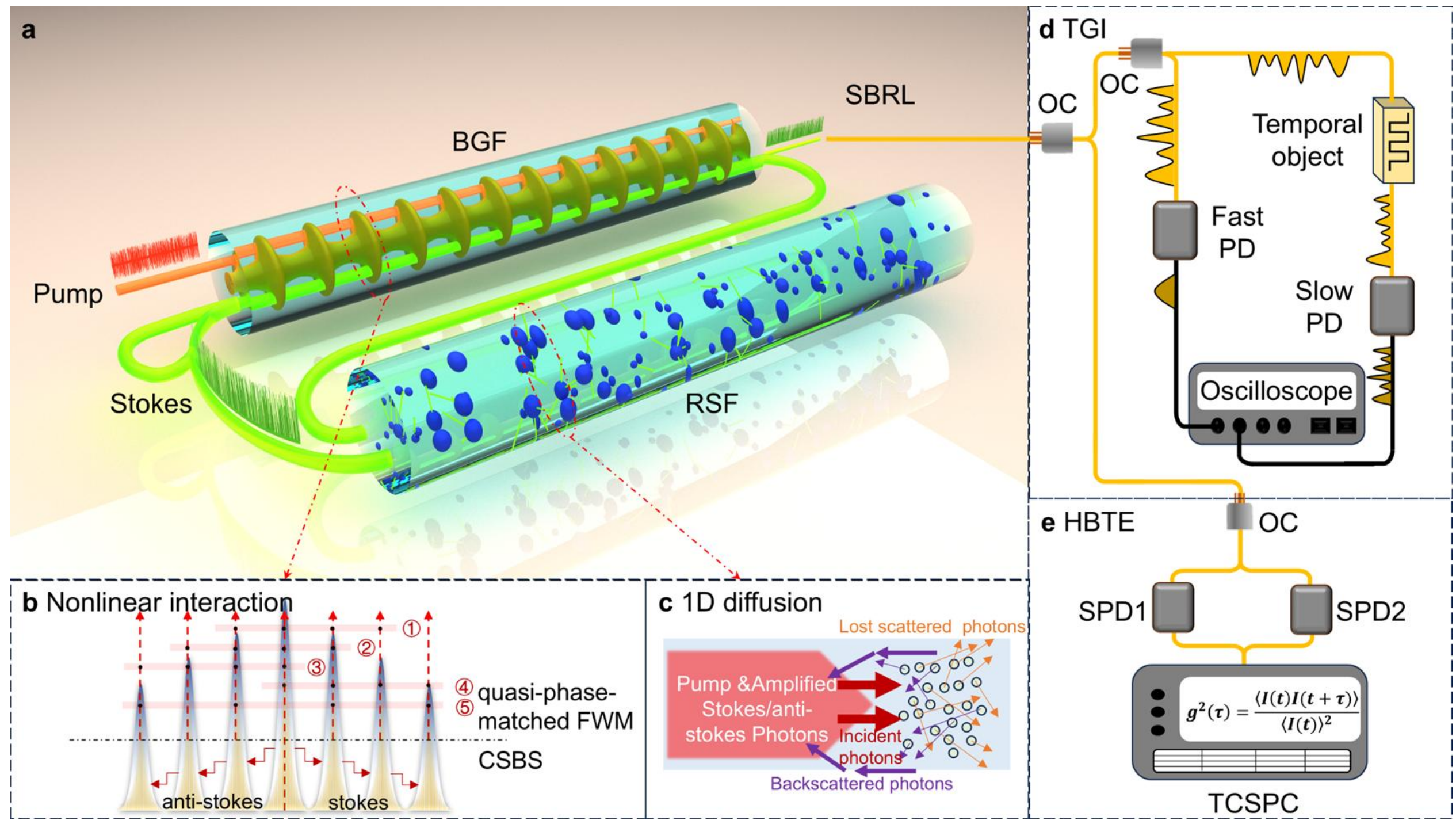

**Fig. 1 | Integrated architecture and operating principles of the superbunched random fiber laser (SRFL), temporal ghost imaging system, and second-order coherence measurement. a**, Schematic of the SRFL composed of a Brillouin gain fiber (BGF) and a Rayleigh scattering fiber (RSF). **b**, Nonlinear interactions inside the BGF, including the CSBS and quasi-phase-matched FWM processes. **c**, One-dimensional random diffusion inside the RSF provided by the Rayleigh scattering process. **d**, Temporal ghost-imaging (TGI) setup. **e**, Hanbury Brown–Twiss experiment (HBTE) setup for measuring second-order coherence. SRFL, superbunched random fiber laser; TGI, temporal ghost imaging; HBTE, Hanbury Brown–Twiss experiment; BGF, Brillouin gain fiber; RSF, Rayleigh scattering fiber; FWM, four-wave mixing; CSBS, cascaded stimulated Brillouin scattering; OC, optical coupler; SPD, single-photon detector; TCSPC, time-correlated single-photon counter; PD, photon detector.

Tailoring the length of the RSF enables precise control over the strength of this random feedback and the baseline level of intracavity intensity fluctuations. Linear wave interference within this disordered fiber section increases the density of indistinguishable photon propagation paths, thereby establishing a strongly fluctuating optical field background that serves as the essential physical foundation for photon bunching. Superbunching becomes particularly pronounced when this random distributed feedback acts in concert with the complex nonlinear dynamics in the BGF **(Fig. 1b)**, which are dominated by the coexistence and mutual coupling of CSBS and quasi-phase-matched FWM. CSBS facilitates the sequential generation of higher-order Stokes components and enables the recursive transfer and amplification of stochastic noise along the nonlinear cascade, while FWM redistributes optical energy and intensity fluctuations among the pump, Stokes, and anti-Stokes fields provided that the local dispersion and polarization conditions within the fiber support efficient phase matching.

Collectively, these coupled linear and nonlinear processes constitute an exceptionally efficient mechanism for amplifying optical field fluctuations in the SRFL. The schematic in Fig. 1b illustrates the stepwise energy transfer from the pump beam to successive higher-order Stokes components, highlighting the progressive accumulation and exponential amplification of stochastic noise across adjacent spectral orders. Each spectral component of the SRFL emission consists of a dense ensemble of narrow resonance modes, which directly reflects the participation of numerous random cavity modes in both the CSBS and FWM processes. The collective involvement of these diverse cavity modes effectively increases the degrees of freedom of the optical system, leading to enhanced stochasticity in the emission field, intricate intermode competition, and a more efficient transfer of intensity fluctuations across the entire emission spectrum.

## Results

### Route to superbunching

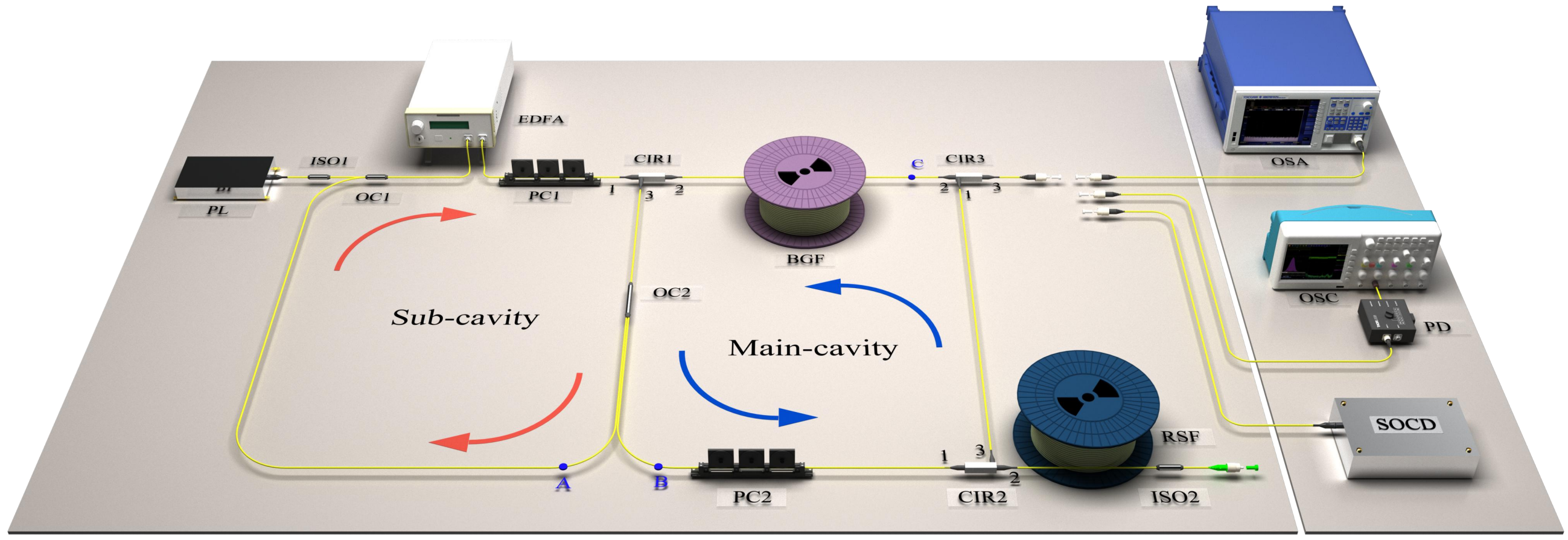

**Fig. 2 | Experimental setup for route to superbunching.** The underlying mechanism was investigated by implementing three sequential operational configurations, with optical processes progressively tailored to unveil the contributions to superbunching. (1) Non-resonant nonlinear scattering regime: Points A, B and C were disconnected, OC2 was removed, and a 5-km single-mode fiber (SMF) served as the BGF, establishing a baseline with negligible nonlinear interactions and linear diffusion. (2) Resonant random lasing regime (single nonlinear process): Point A was disconnected, points B and C were connected, OC2 was removed; a 5-km SMF and a 10-km SMF were employed as the BGF and the RSF, respectively. This configuration enabled SBS with random distributed

feedback, but no CSBS or FWM was initiated. (3) Resonant random lasing regime (coupled nonlinear processes): All points A, B and C were connected, OC2 was inserted; a 390-m HNLF and a 10-km SMF acted as the BGF and RSF, respectively. This full SRFL configuration realized random distributed feedback with the co-initiation of CSBS and quasi-phase-matched FWM.

To disentangle the individual contributions of linear diffusion and nonlinear interactions to the enhancement of $g^{(2)}(0)$, we implemented the three configurations depicted in **Fig. 2** and measured the zero-delay second-order coherence in each case. The evolution of $g^{(2)}(0)$ shown in **Fig. 3** traces a continuous pathway, from nonresonant spontaneous scattering, to random lasing with distributed feedback, and finally to robustly superbunched emission.

In the first configuration, points A, B and C were left unconnected and OC2 was removed, with a 5-km standard single-mode fiber (SMF) serving as the BGF. This cavity configuration cannot sustain resonant modes, precluding the development of CSBS and FWM, and eliminating distributed feedback. As shown in **Fig. 3a**, $g^{(2)}(0)$ first rises above 1, exceeds 2 near the Brillouin threshold, and subsequently falls back toward 1. Well below threshold, Stokes emission is thermally seeded, yet the detected photon flux is sufficiently low that shot noise dominates the statistics. Near threshold, intensity fluctuations are intermittently amplified into short bursts that induce pump depletion, generating high-intensity spikes and yielding $g^{(2)}(0)>2$ [29]. Far above threshold, SBS forms a quasi-steady-state Stokes wave; combined with gain saturation, pump depletion suppresses noise growth, driving the photon statistics toward the Poisson limit. This behavior marks the crossover from a fluctuation-dominated spontaneous scattering regime to a stable stimulated emission regime in a nonresonant cavity.

In the second configuration, points B and C were connected and the RSF was incorporated to introduce distributed feedback and linear diffusion. As illustrated in **Fig. 3b**, $g^{(2)}(0)$ initiates at values well above those in the first configuration and decreases monotonically with increasing pump power. This enhanced initial photon bunching arises from spontaneous Stokes photons exploring numerous scattering loops and retracing multiple propagation paths within the cavity, which amplifies intensity fluctuations via wave interference even at low pump power. A transient increase in $g^{(2)}(0)$ near the threshold again reflects instability-driven flickering[29]. At higher pump powers, random lasing modes develop, with positive feedback selectively amplifying only those propagation paths with maximum gain. This mode selection, accompanied by gain saturation and pump depletion, suppresses large intensity excursions and reduces relative noise, pushing the photon statistics toward a Poisson-like regime. In comparison with **Fig. 3a**, this configuration demonstrates that linear diffusion alone tends to improve second-order coherence once random lasing is established, provided that strong nonlinear coupling is absent.

The third configuration, corresponding to the full SRFL, enables the coexistence of linear diffusion and cascaded nonlinear interactions, leading to a stark qualitative change in emission behavior. As shown in **Fig. 3c**, $g^{(2)}(0)$ for the 1st, 4th and 8th Stokes orders, as well as the 2nd, 5th and 6th anti-Stokes orders, increases rapidly with pump power. Higher-order spectral components exhibit a steeper rise and reach substantially larger $g^{(2)}(0)$ values than lower-order ones, indicating a transition to a nonlinearity-dominated emission regime. For the present configuration and pump power range, the 8th Stokes component achieves a maximum $g^{(2)}(0)$ of approximately 26. The RSF continues to provide a dense set of indistinguishable propagation paths that seed intensity fluctuations, while CSBS and quasi-phase-matched FWM synergistically redistribute energy and noise across the entire spectral comb. Each higher-order component inherits and further amplifies the fluctuations of the preceding order, resulting in intensity profiles characterized by rare yet extremely large amplitude spikes. This causes the second-order intensity moment to grow far more rapidly than the mean intensity, yielding the exceptionally high $g^{(2)}(0)$ values observed. A direct comparison of **Fig. 3b** and **Fig. 3c** reveals that once cascaded nonlinear coupling is activated, linear diffusion no longer stabilizes the emission; instead, it acts as a robust driver of intensity fluctuations. This synergetic mechanism, which couples linear diffusion with cascaded nonlinearities, fundamentally transforms the random fiber laser into a fluctuation-dominated source of superbunched light.

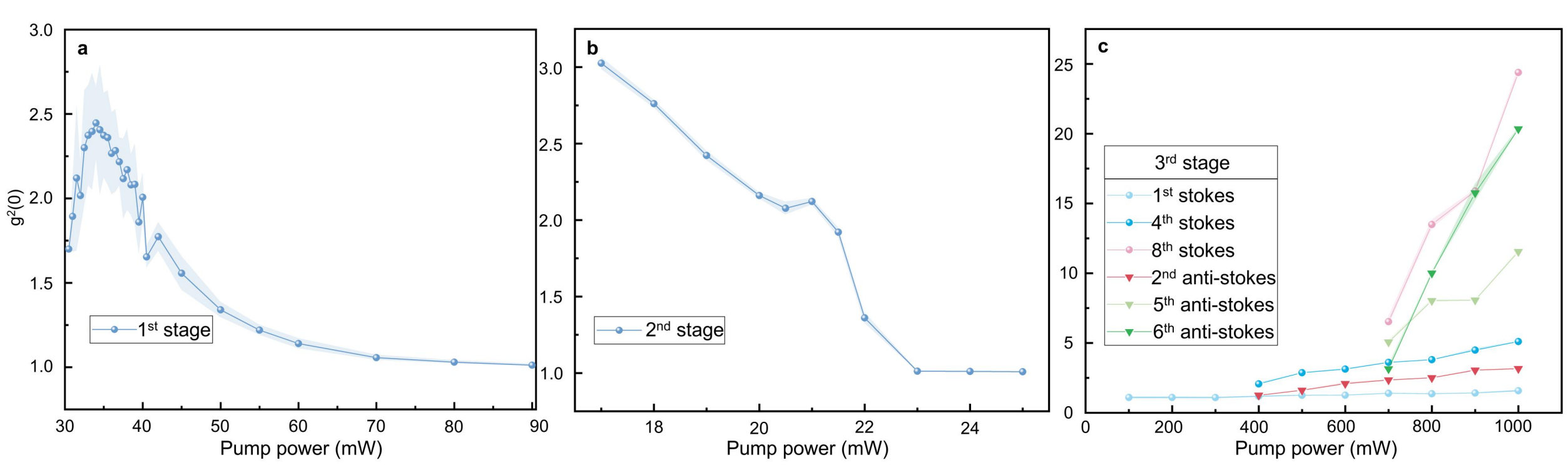

**Fig. 3 | Evolution of zero-delay second-order coherence with pump power under different operational regimes. a**, Non-resonant nonlinear scattering regime. **b**, Resonant random laser regime with linear diffusion and single nonlinear scattering process. **c**, Resonant random laser regime with linear diffusion and coupled nonlinear interactions.

## In-depth characterization of second-order coherence in the SRFL

The operating principle outlined above indicates that higher-order optical components should exhibit more pronounced statistical fluctuations. This hierarchical trend is experimentally verified in **Fig. 4**. Panel **Fig. 4a** shows the full emission spectrum, featuring a dense comb structure composed of eight Stokes and eight anti-Stokes orders. This distinct spectral signature confirms that the laser operates in a regime dominated by strong CSBS and FWM. To characterize the intrinsic properties of individual spectral components, we performed spectral filtering with a tunable optical filter. The filtered spectra in **Fig. 4 b1-b4** display narrow, well-resolved and fully isolated peaks for the 1$^{st}$ and 8$^{th}$ Stokes orders and for the 1$^{st}$ and 6$^{th}$ anti-Stokes orders, ensuring spectral crosstalk is eliminated and the accuracy of subsequent temporal and statistical analyses is guaranteed. The corresponding temporal intensity traces in **Fig. 4 c1-c4** clearly reveal a monotonic increase in fluctuation intensity with the order of Stokes/anti-Stokes components. The 1$^{st}$ Stokes and anti-Stokes components exhibit moderate, relatively regular fluctuations, whereas the 8$^{th}$ Stokes and 6$^{th}$ anti-Stokes components show intense, irregular burst-like dynamical behavior, a direct manifestation of cumulative noise amplification in higher-order nonlinear cascades.

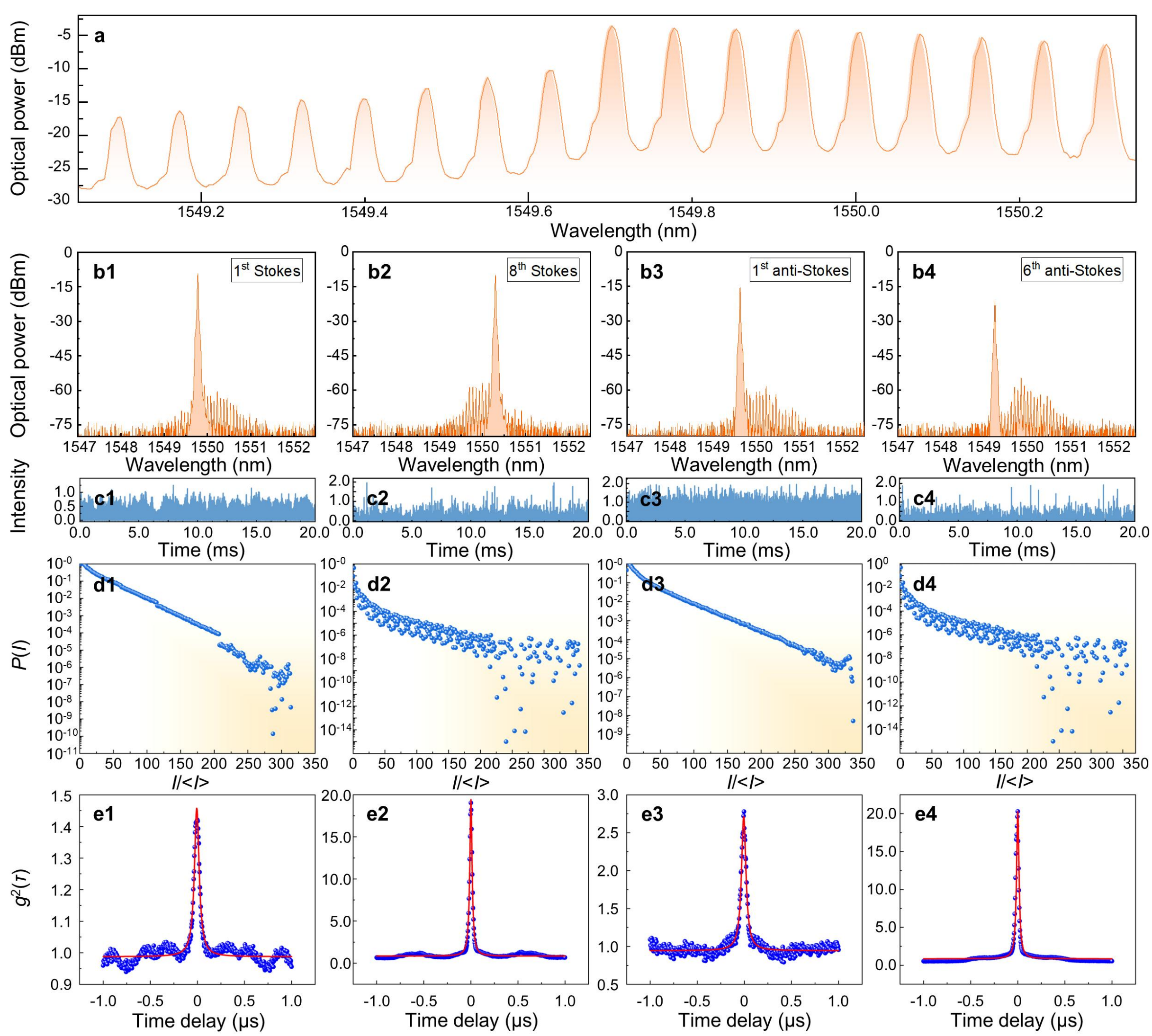

**Fig. 4 | Spectral characteristics, temporal dynamics, statistical distributions and second-order coherence of representative Stokes and anti-Stokes components in the SRFL. a**, **Full** optical spectrum of the SRFL with ten orders of Stokes and eight orders of anti-Stokes components. **b1–b4**, Filtered optical spectra of the 1st-order and 8th-order Stokes emissions, and the 1st-order and 6th-order anti-Stokes emissions. **c1–c4**, Corresponding temporal intensity traces of the filtered spectral lines in b1–b4, revealing the evolution of fluctuation strength with spectral order. **d1–d4**, Normalized intensity probability distributions of the temporal traces in c1–c4, quantifying the statistical deviation from quasi-thermal behavior. **e1–e4**, Measured second-order coherence functions $g^{(2)}(\tau)$ for the corresponding Stokes/anti-Stokes components, with the zero-delay value $g^{(2)}(0)$ quantifying photon bunching strength.

This escalating trend of fluctuations with spectral order is further quantified and validated by normalized intensity distributions and second-order coherence function. As shown in **Fig. 4 d1**, the 1st Stokes component exhibits an almost exponential distribution with a relatively short tail, a hallmark characteristic of quasi-thermal light with weak intensity fluctuations. In contrast, the distributions for the 8th Stokes and 6th anti-Stokes components (**Fig. 4 d2–d4**) span multiple orders of magnitude in intensity and exhibit prominent heavy-tailed profiles, which signify the occurrence of rare yet extremely intense intensity bursts arising from the cumulative amplification of stochastic fluctuations across the nonlinear cascade. Consistent with these intensity distribution characteristics, the temporal $g^{(2)}(\tau)$ curves in **Fig. 4 e1-e4** show that the1st Stokes component exhibits only modest bunching, with $g^{(2)}(0)$ slightly below the thermal value of 2. By contrast, the higher-order components exhibit sharp, prominent peaks at zero delay with $g^{(2)}(0)$ far exceeding 2. This pronounced superbunching behavior unequivocally confirms that multi-stage noise transfer and amplification within the nonlinear cascade constitute the fundamental physical origin of the observed extreme photon correlations.

To quantitatively and systematically characterize the second-order coherence of the SRFL, we mapped the zero-delay second-order coherence $g^{(2)}(0)$ as a two-dimensional parametric landscape governed by pump power, spectral order, and RSF length. As illustrated in the parametric maps for a fixed 5-km RSF (**Figs. 5a, b**), all spectral components remain in a near-coherent emission regime with strongly suppressed intensity fluctuations at low pump powers. However, once the pump power exceeds the threshold for activating deep nonlinear cascades, $g^{(2)}(0)$ rises steeply, especially for higher-order components. A distinct region of giant superbunching emerges in the high-power, high-order domain, confirming that the synergy between long nonlinear cascade chains and strong pumping is indispensable for pronounced fluctuation amplification. In support of this conclusion, extending the RSF length to 15 km significantly enhances the achievable superbunching strength (**Figs. 5c, d**). This enhancement stems from the increased density of indistinguishable optical propagation paths provided by the longer scattering medium, which creates a more abundant stochastic background to seed and drive the subsequent nonlinear interaction cascade.

To isolate and quantify the unique role of the linear diffusion mechanism, we performed a statistical analysis of coherence for various RSF lengths while maintaining constant spectral laser efficiency, and compared the results with the mirror-based feedback control configuration. Mirror-feedback measurements yield $g^{(2)}(0)$ values consistently lower than those obtained with a 1 km RSF (**Figs. 5e, f**). Furthermore, $g^{(2)}(0)$ increases monotonically across all spectral orders as the fiber length is increased from 1 km to 25 km, and the maximum second-order coherence systematically shifts toward higher spectral orders. These results unambiguously confirm that both the magnitude and the spectral range of superbunching are governed by the synergistic interplay between cascaded nonlinearities and the stochastic fluctuation seeds provided by linear diffusion.

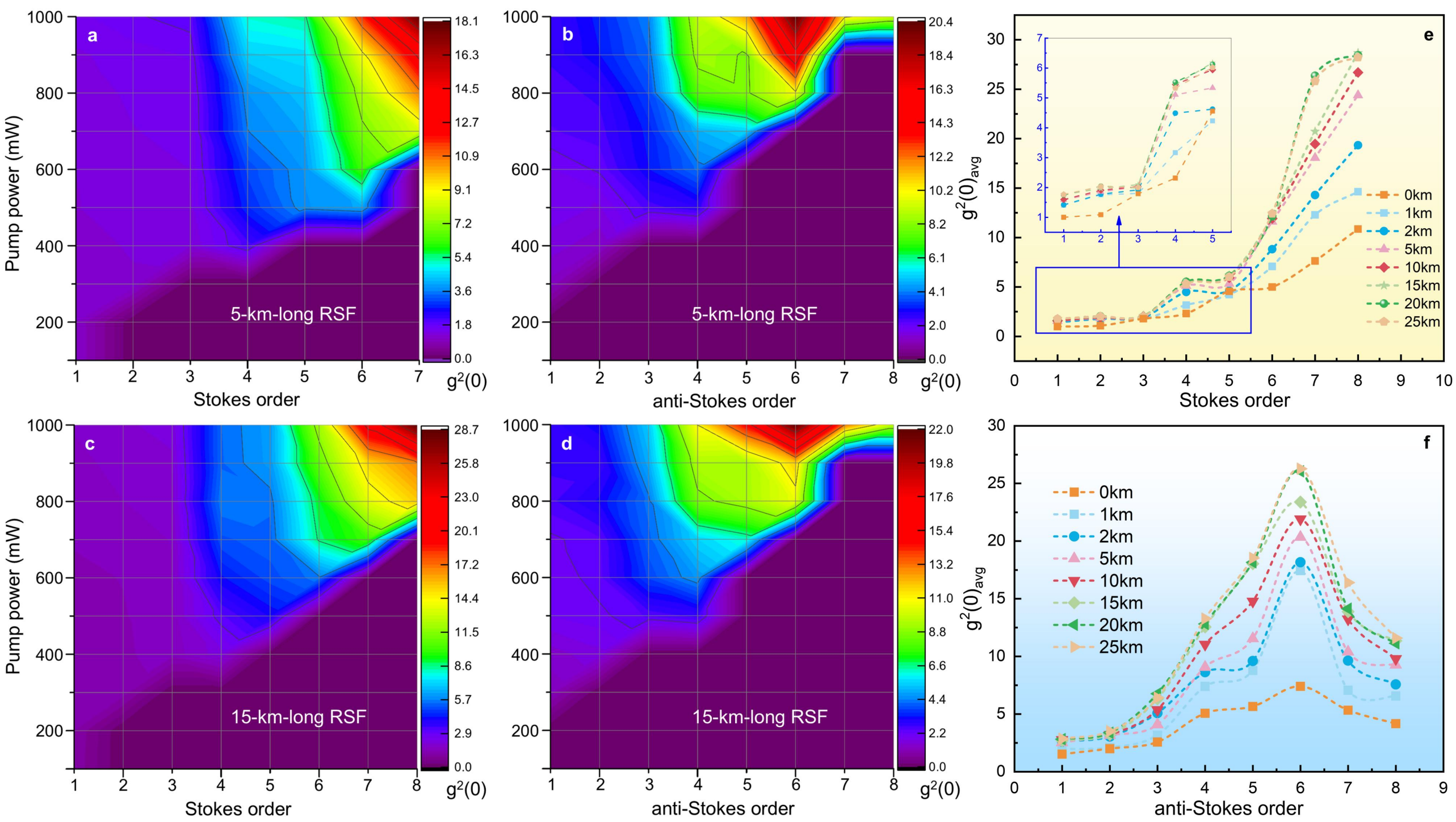

**Fig. 5 | Parameter-dependent landscape of second-order coherence in the SRFL.** Two-dimensional color maps of the measured zero-delay second-order coherence $g^{(2)}(0)$ of Stokes and anti-Stokes components as functions of pump power at fixed RSF lengths of 5 km (**a**,**b**) and 15km (**c**,**d**). **e**,**f**, Zero-delay second-order coherence $g^{(2)}(0)$ versus spectral order for various RSF lengths, with the spectral laser efficiency held constant across all measurements.

Building upon the parametric characterization of superbunching, we investigate the intrinsic physical link between these extreme photon statistics and the macroscopic phase structure of the SRFL. Using the statistical framework of spin glasses to analyze the laser emission, we characterize photonic phase behaviors using the mean Parisi overlap parameter $|q|_{average}$, while quantifying the microscopic photon statistics via the zero-delay second-order coherence $g^{(2)}(0)$. As illustrated in **Figs. 6a and 6b,** these two physical quantities exhibit a striking anticorrelation across all examined Stokes and anti-Stokes orders. Lower-order components are characterized by large $|q|_{average}$ and weak photon bunching behavior. In contrast, higher-order components evolve toward a non-equilibrium paramagnetic regime, where $|q|_{average}$ approaches zero and photon bunching is dramatically enhanced.

In the low-order regime, strong replica correlations indicate a frozen energy landscape that restricts the photon statistics to a near-thermal state. As the spectral order increases, this glassy order gradually dissolves and the system is driven into a disordered paramagnetic phase—a transition marked by the decay of $|q|_{average}$ toward zero. This disappearance of macroscopic glassy order effectively releases the constraints imposed on the optical field, enabling the emergence of giant superbunching with exceptionally strong intensity correlations. This physical scenario is further validated by the pump-power-dependent results shown in **Figs. 6c–6h,** where increasing the pump power drives the system from a glassy regime with high $|q|_{average}$ and low $g^{(2)}(0)$ into a paramagnetic state with reduced overlap and significantly enhanced $g^{(2)}(0)$ Collectively, these results demonstrate that the glassy-to-paramagnetic phase transition serves as a macroscopic indicator for the onset of giant superbunching, establishing a rigorous bridge between fundamental concepts in complex disordered systems and measurable microscopic photon statistics.

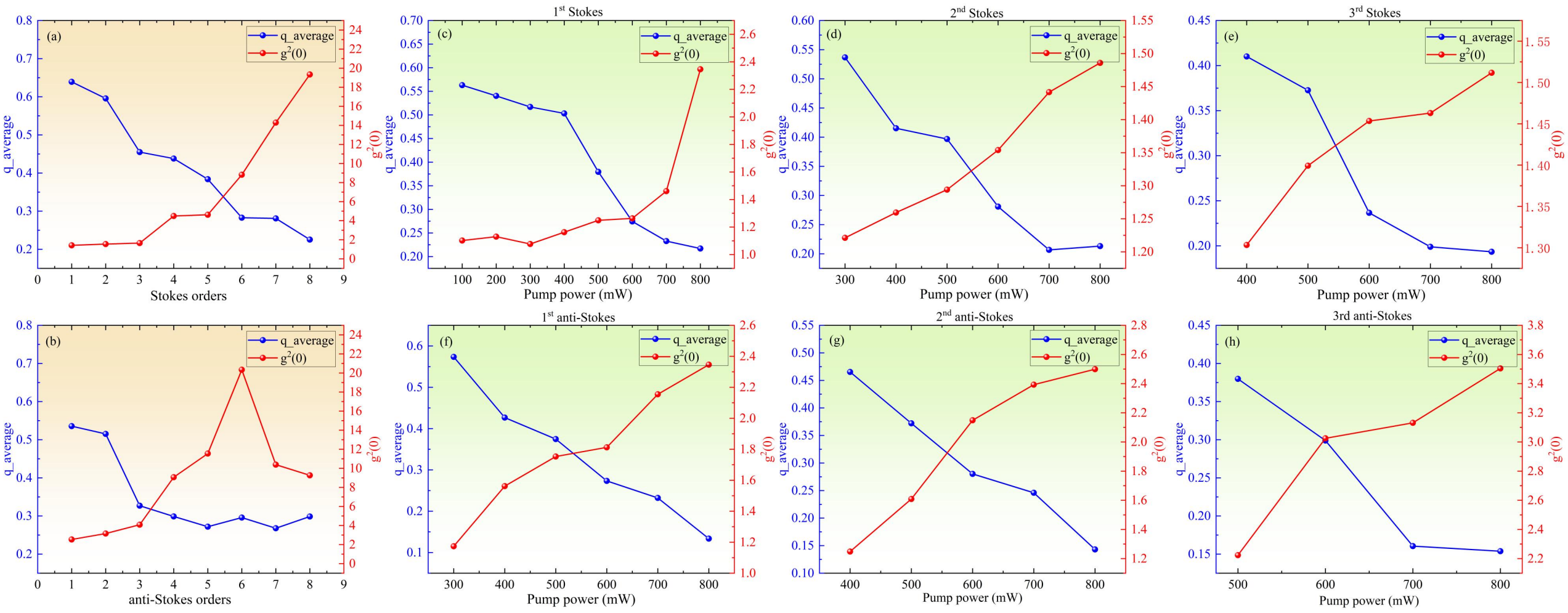

**Fig. 6 | Correlation between photonic phase transitions and photon-counting statistics in the SRFL. a**, **b**, Evolution of the mean Parisi overlap parameter $|q|_{average}$ (blue) and zero-delay second-order coherence $g^{(2)}(0)$ (red) versus Stokes and anti-Stokes order,

respectively. The data reveal a clear inverse relationship: increasing disorder (lower $|q|_{average}$) corresponds to stronger superbunching. **c**–**h**, Evolution of $|q|_{average}$ and $g^{(2)}(0)$ as functions of pump power for the 1st, 2nd, and 3rd-order Stokes (**c**–**e**) and anti-Stokes (**f**–**h**) components. These results demonstrate that the spin-glass-to-nonequilibrium-paramagnetic phase transition drives the emergence of extreme photon correlations.

## Temporal ghost imaging with SRFL

Following the characterization of tunable superbunching, we assess the practical applicability of the SRFL in a correlation-based optical detection architecture. Specifically, high-$g^{(2)}(0)$ Stokes components are utilized as structured illumination for temporal ghost imaging (TGI). In this configuration, the temporal profile of a target object is reconstructed via intensity correlations rather than direct intensity detection, making the reconstruction fidelity inherently dependent on the second-order coherence of the illumination source. Representative Stokes and anti-Stokes orders were spectrally filtered using a bandwidth-tunable optical filter and injected into the TGI setup (**Fig. 1d**). To ensure that the observed imaging fidelity predominantly reflects the intrinsic statistical properties of the source, rather than artifacts from excessive ensemble averaging, we fixed the number of realizations to $N = 10{,}000$ for all measurements.

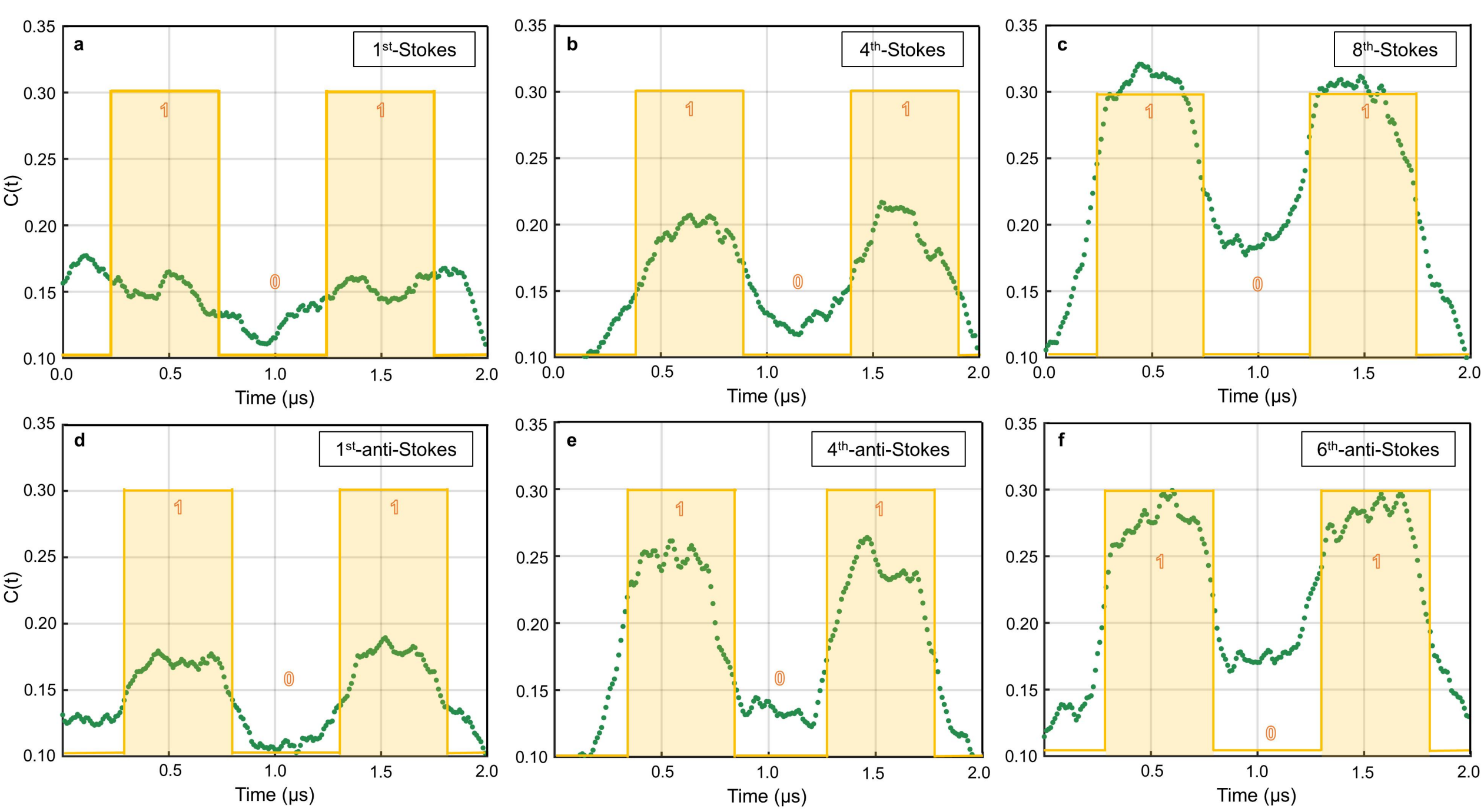

**Fig. 7 | Temporal ghost-imaging reconstructions using Stokes and anti-Stokes components of different spectral orders.** Reconstructed temporal ghost images are plotted as green dashed curves. All correlation reconstructions use a fixed number of realizations: $N = 10{,}000$. Yellow shaded regions and the overlaid 0/1 labels denote the temporal binary object encoded in the object arm. Panels a–c illustrates reconstructions using 1st-, 4th- and 8th-order Stokes components, and Panels d–f illustrates reconstructions

using 1st-, 4th- and 6th-order anti-Stokes components.

The reconstruction performance for different spectral orders is summarized in **Fig. 7**. The retrieved temporal waveforms show a striking contrast in fidelity between low-order and high-order components. Reconstructions based on low-order components exhibit severe background noise and low imaging contrast, consistent with their thermal-like or quasi-coherent photon statistics. In sharp contrast, high-order Stokes and anti-Stokes components yield clean, well-defined temporal profiles with drastically improved signal-to-noise ratios (SNRs). This dramatic performance improvement originates directly from the giant superbunching enabled by the synergistic interplay of linear diffusion and cascaded nonlinearities, which generates the strong, high-contrast intensity fluctuations essential for high-efficiency TGI. Additional control measurements further reveal that even doubling the number of averages to $N = 20{,}000$ remains insufficient to recover a clear object trace for low-order components, whereas high-order components already achieve high-fidelity reconstruction with only $N = 10{,}000$ averages. These results unambiguously demonstrate that engineering second-order coherence via controlled disorder and nonlinear coupling offers a powerful route toward high-performance TGI. By optimizing the underlying photon statistics instead of merely increasing optical power or acquisition time, the SRFL effectively mitigates the fundamental trade-off between imaging speed and SNR in conventional correlation-based sensing platforms.

## Discussions

The realization of tunable photon superbunching in a fully fiber-integrated architecture represents a major advance over conventional strategies for generating super-thermal light fields. Conventional pseudo-thermal sources, typified by rotating ground-glass diffusers, rely on mechanical modulation, which fundamentally limits their operational bandwidth and integrability. In contrast, our SRFL harnesses the intrinsic complexity of linear Rayleigh scattering and nonlinear optical gain to generate substantial intensity fluctuations without any moving parts. This platform enables a robust, all-optical strategy for engineering tailored photon statistics and achieves superbunching behavior that greatly exceeds the Gaussian-thermal limit. Consequently, it supports correlation-contrast levels unattainable using conventional chaotic emitters under equivalent experimental conditions.

This remarkable improvement in statistical photon properties carries important implications for correlation-based optical sensing technologies. As revealed by our comparative ghost imaging experiments, the striking difference in imaging performance between the superbunched high-order output and low-order thermal-like components signifies a substantial advance in practical functionality. Whereas standard thermal sources fail to rise above the noise floor even with extensive ensemble averaging, the SRFL enables high-fidelity reconstruction with drastically reduced data volumes. This suggests that the heavy-tailed statistical nature of the SRFL emission naturally enhances correlation signals, thereby substantially reducing the acquisition time and computational overhead required for high-performance ghost imaging. This attribute is particularly critical for dynamic sensing scenarios that demand high frame rates and real-time data processing.

Beyond TGI, the ability to control and tailor photon statistics through engineered disorder establishes the SRFL as a highly versatile photonic platform for diverse applications. The rich interaction between linear diffusion and cascaded nonlinear processes forges a conceptual bridge between classical random lasing and photon-counting statistics widely investigated in quantum optics. This connection provides new fundamental insights into the coherence properties of disordered photonic systems. Looking forward, this easily implementable, high-spectral-density source holds significant promise for applications that demand high-entropy optical illumination, including secure optical communications, fluctuation-enhanced LiDAR, and non-invasive biological sensing within strongly scattering biological environments.

## Acknowledgements

This work was financially supported by Qingdao Municipal Natural Science Foundation (25-1-1-167-zyyd-jch), Natural Science Foundation of Shandong Province (ZR2025MS1048), National Grant Program for High-level Returning Oversea Talents, National Natural Science Foundation of China (62105180), Taishan Scholar Foundation of Shandong Province (tsqn202211027) and Qilu Young Scholar Program of Shandong University.

**Author Contributions.** Q.J. and Y.P. performed the experiments, analyzed the data, developed the theoretical model, and wrote the manuscript. W.Z. assisted with the experiments and data analysis. X.Z. provided experimental resources and technical support. Z.Q. contributed to data interpretation and manuscript discussion. Z.L. contributed to data interpretation, manuscript discussion, and supervised the project. Y.X. conceived the idea, supervised the project, designed the research framework, revised the manuscript, and secured funding. All authors read and approved the final manuscript.

**Conflict of interest.** Authors declare that they have no competing interests.

**Data and materials availability.** All data are available in the main text or the supplementary materials